\documentclass{cpcauth}

\usepackage{amssymb}

\begin{document}

\begin{frontmatter}



\title{Fortran MPI Checkerboard Code for SU(3) Lattice Gauge Theory I} 


\author{Bernd A. Berg$^{\,a}$}

\address{~~\\
$^{a)}$ Department of Physics, Florida State University, 
Tallahassee, FL 32306-4350, USA}

\date{\today} 

\begin{abstract}
We document Fortran MPI checkerboard code for Markov Chain Monte Carlo 
simulations of pure SU(3) lattice gauge theory with the Wilson action
on a D-dimensional double-layered torus. This includes the usual torus
with periodic boundary conditions as an optional case. We use 
Cabibbo-Marinari heatbath checkerboard updating. Parallelization 
on sublattices is implemented in all D directions and can be restricted 
to less than D directions. The parallelization techniques of this 
paper can be used for any model with interactions of link variables
defined on plaquettes.
\medskip

\noindent {\bf Program Summary} \smallskip

\noindent {\it Program title: \tt STMC2LSU3MPI.}

\noindent {\it Program identifier:} Not yet available.

\noindent {\it Program summary URL:} Not yet available.

\noindent {\it Program available from:} Temporarily from URL
{\tt http://www.hep.fsu.edu/\~\,$\!$berg/research}

\noindent {\it Programing language:} Fortran~77 with MPI extensions.

\noindent {\it Computer:} Any capable of compiling and executing 
Fortran~77 code with MPI extensions.

\end{abstract}


\end{frontmatter}

\section{Introduction} \label{sec_intro}

Moore's law \cite{Mo65} appears to be dead. Certainly we have not seen
CPU processor speed going up by a factor of ten in the last five years.
Instead, we get now ten times as many processors (more precisely cores) 
for the price of one five years ago. PCs with 8~cores have become 
commodities and soon one may expect 64 or more. The usefulness of 
parallelization is no longer limited to large scale supercomputer 
applications, but becomes relevant for everyday calculations.

This motivates the present paper, which documents Fortran~77 MPI 
checkerboard \cite{BaMo82} code for Markov Chain Monte Carlo (MCMC) 
simulations of pure SU(3) Lattice Gauge Theory (LGT) with the Wilson 
action on D-dimensional lattices.  Sublattices are updated in parallel 
after collecting boundary variables from other sublattices. The 
introduced parallelization techniques apply to any model with 
dynamical variables defined on links and their interactions on
plaquettes.

The code of this paper
implements the Cabibbo-Marinari (CM) SU(3) updating \cite{CaMa82} 
using for the SU(2) subgroups the heatbath method of Fabricius-Haan 
\cite{FaHa84} and Kennedy-Pendleton \cite{KePe85} (FHKP), which is
more efficient than the older Creutz heatbath \cite{Cr80}. CM with 
FHKP SU(2) updating is also about three times more efficient than 
Pietarinen's \cite{Pi81} full SU(3) heatbath \cite{Ba07}. 

To synchronize the simulations on all processes, we use FHKP 
updating in the multi-hit accept/reject version~\cite{FrMa87}. 
Overrelaxation moves \cite{Ad88} are presently not implemented, but 
would fit seamlessly into the code. In extension of the usual periodic 
boundary conditions (PBC), which define the gauge system on a torus, 
our code allows for a double-layered torus (DLT). These are two 
identical lattices, each using the other as boundary, a geometry 
expected to be of relevance for studies of the deconfining phase 
transition.

The next section gives an overview of the code and explains Web access. 
Section~\ref{Verify} provides a number of verifications. Summary and 
conclusions follow in section~\ref{Conclusions}. Runs are setup in 
the code, which reproduce the examples of this and a companion 
paper~\cite{BeWu09}. Running on up to 1$\,$296 CPU cores, the 
companion paper studies performance as function of the number of 
MPI processes. It also discusses and resolves problems, which were 
encountered with MPI send and receive instructions for large arrays. 

\section{Overview of the Code} \label{sec_code}

The code for this paper is freely available as a gzipped archive 
$$ {\tt STMC2LSU3MPI.tgz} $$
that can be downloaded from the website of the author 
\smallskip

\centerline{\tt http://www.hep.fsu.edu/\~\,$\!$berg/research\ .}
\smallskip

\noindent With \smallskip

\begin{figure}[t]
 \begin{picture}(150,155)
    \put(0, 0){\includegraphics{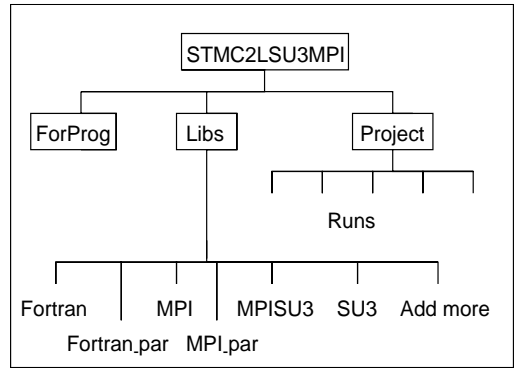}}
  \end{picture}
\caption{Structure of our program package.\label{fig_STMC2LSU3}}
\end{figure}

\centerline{\tt tar -zxvf STMC2LSU3MPI.tgz} \smallskip

\noindent the folder structure of Fig.~\ref{fig_STMC2LSU3} is created. 
Main programs are located in {\tt ForProg}. The {\tt LIBS} folder
contains a number of libraries with plain Fortran~77 and Fortran~77 
MPI code. Test and verification runs are setup in subfolders of
several {\tt Project} folders. Non-MPI SU(3) code and runs are
in the project tree {\tt STMCSU3}.

We use checkerboard labeling \cite{BaMo82} to divide lattice sites
into two sets of colors $i_c=1,\,2$. Moving one step in any direction 
changes the color. For spin models with nearest neighbor interactions 
this allows one to update spins at half of the sites in parallel. For 
SU(3) LGT the matrices are located on lattice links and one can update 
at half of the sites one of the link directions in parallel. This 
is employed to update sublattices in parallel after collecting 
from other sublattices boundary variables, which need no updating 
because they belong to another checkerboard or link direction. For 
efficient performance the sublattice volume to surface ratio, each 
measured in numbers of variables, has to be sufficiently large. 
Examples are discussed in~\cite{BeWu09}.

To arrange storage of our SU(3) matrices and other physical variables,
we label lattice sites and links following the book of the 
author~\cite{Be04}. Corresponding routines from {\tt ForLib} of this 
reference are taken 
over into {\tt Libs/Fortran} of Fig.~\ref{fig_STMC2LSU3}. In
our approach a lattice site is specified by a single integer $i_s$, 
which we call {\it site number}. The dimension of the lattice is given 
by D. The Cartesian coordinates of a site are chosen to be
\begin{equation} \label{site_coordinate}
  x^i = 0, \dots ,n^i-1 ~~~{\rm for}~~~ i=1, \dots , D.
\end{equation}
The site number is defined by the formula
\begin{equation} \label{i_s}
  i_s = 1 + \sum_{i=1}^D x^i\,n_a^i,\
  n_a^i = \cases{ 1~~{\rm for}~~i=1, \cr
  \prod_{j=1}^{i-1} n^j~~{\rm for}~~i>1,}
\end{equation}
and calculated by the Fortran function {\tt isfun.f}. Vice versa, 
the coordinates for a given site number $i_s$ are obtained by an 
iteration procedure, which relies on Fortran integer division ({\it 
i.e.}, $1=[5/3]=[5/5]$, $0=[4/5]$, etc.). Let $n_s=\prod_{i=1}^D n^i$ 
be the number of lattice sites. Then,
\begin{equation} \label{xD}
  x^D = \left[{(i_s-1)\over (n_s\,/\,n^D)}\right]
      = \left[{(i_s-1)\over n_a^D}\right]
\end{equation}
and for $i = D-1,\dots ,1$
\begin{equation} \label{x^i}
  x^i = [(j_s^i-1)\,/\,n_a^i)],~~
  j_s^i=i_s-\sum_{j=D}^{i+1} x^j\,n_a^j\,.
\end{equation}
The Fortran subroutine {\tt ixcor.f} computes coordinates from the site 
number, though somewhat differently than by the formulas written down 
here.

The site number $i_s$ allows one to store variables at sites in 1D 
arrays $A_1(n_s)$, independently of the lattice dimension D. Variables 
on links are located in 2D arrays $A_2(n_s,nd)$, where the integer 
$nd$ is the lattice dimension D, $nd\ge 2$ for LGT. One more label 
is required to store SU(3) matrix elements in a 3D array. For 
checkerboard labeling we arrange the lattice variables in two arrays, 
corresponding to the colors $i_c=1,\, 2$. The formula returning the 
color assignment of a lattice site is
\begin{equation} \label{i_c}
  i_c = 1 + {\rm mod}\left[\sum_{i=1}^D x^i,2\right]\ .
\end{equation}
To update variables in array~1, neighbor variables are collected from 
array~2, which remains unchanged, and vice versa. LGT requires also to
collect variables from the same checkerboard, which are not updated, 
because they are on links in other directions than the one updated. 
The checkerboard algorithm requires even lattice extensions. Otherwise 
PBC destroy the pattern.

\subsection{Updating}

Our code implements CM \cite{CaMa82} SU(3) updating using for the 
SU(2) subgroups the FHKP \cite{FaHa84,KePe85} heatbath algorithm. In 
the original FHKP version proposals are repeated until one is accepted, 
which is by construction from the desired probability distribution. For 
parallelization this is inconvenient, because all MPI processes have 
to wait until the last one finished. As pointed out by Fredenhagen and 
Marcu~\cite{FrMa87}, one can terminate the inner loop after a finite 
number of hits and keep the link matrix at hand when none of the 
proposal has been accepted. The new configuration is still proposed 
with the local heatbath distribution. What changes is the average 
stay time of the old configuration. This time depends on the 
configuration at hand, but drops out in the detailed balance equation. 
By using CM heatbath in this Metropolis-like fashion the MPI processes 
get synchronized. The 1-hit acceptance rate depends on $\beta$ and is 
in the scaling region of SU(3) LGT around 97\%. Lower 1-hit acceptance 
rates are encountered for smaller $\beta$ values. One may then 
increase the number of hits. 

As usual, the updating step keeps track of the total action. Due 
to parallelization action differences have to be added by the MPI
process of the sublattice on which the update is carried out. Then,
{\it action fluctuation across boundaries} can be created, which 
lead in course of time to absurd sublattice contributions, while the 
total action (their sum) is still correct. To elaborate on this point, 
we first need to define sublattice actions. Due to links crossing 
boundaries, there is some amount of freedom in that. We simply 
attribute the action of a plaquette to the sublattice, which contains 
the site from which two forward links of the plaquette emerge. Now, 
updating one of the other links of a plaquette, the action change is 
recorded in a wrong sublattice, if the updated link emerges there. 

As long as one is only interested in the total action, it is sufficient 
to recalculate the sublattice action once in a while directly to prevent 
an amplification of rounding errors due to differences of large numbers 
(sublattice contributions can fluctuate to negative values). However, 
if one wants to attribute physical significance to sublattice actions, 
it is mandatory to recalculate them before a measurement is recorded.

Our updating subroutine is {\tt cbsu3\_2hbnhit.f} located in {\tt
Libs/MPISU3}. A call to this routine performs one sweep, which is here 
defined by updating each SU(3) matrix once in {\it sequential} order 
(see section~\ref{subsec_code} for more details). Updating in sequential 
order fulfills balance and is more efficient than updating link matrices
in random order. This observation holds also for spin models~\cite{Be04}.

\subsection{Double-Layered Torus}

This section can be skipped by readers, who are only interested in 
simulations with PBC. The DLT is, for instance, of interest for 
simulations on $N_s^3\,N_{\tau}$ lattices if one likes to have 
boundaries at a different temperature than the interior of the 
lattice~\cite{BaBe07}, as it is the case for deconfined volumes created 
in relativistic heavy ion collisions. The DLT is defined by two lattices 
of identical size, each using the other as boundary in all or just
volume directions. In the latter case distinct $\beta$ values in the
lattices lead to different physical temperatures $T$ through the 
usual definition of $T=1/(aN_{\tau})$.

\begin{figure}[t]
 \begin{picture}(150,155)
    \put(0, 0){\includegraphics{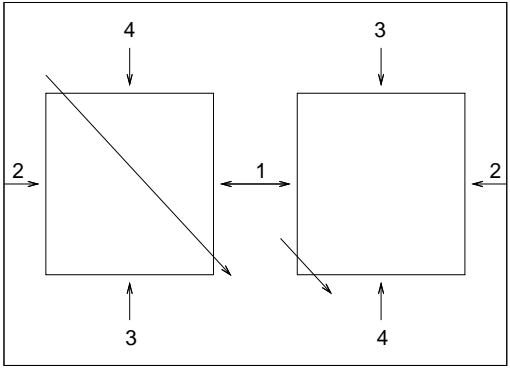}}
  \end{picture}
\caption{2D Double layered torus. \label{fig_DLT}}
\end{figure} 

Even with identical $\beta$ values in both lattices the DLT has some 
intriguing properties as illustrated in Fig.~\ref{fig_DLT} for a 2D 
DLT of size $(N_s)^2$. The boundaries are glued together as indicated 
by the arrows\footnote{Note that interchanging the labels 3 and~4 on 
one of the lattices of the figure leads to an undesirable situation 
in which some sites pairs are connected by two links}. 
While for PBC the shortest connection of a point with itself through 
the boundary is of length $N_s$, it is now of length $\sqrt{2}\,N_s$ 
along the diagonal. The two arrows in diagonal direction give an 
example of a line, which is closed by DLT boundary conditions. Compared 
to a torus of size $(N_s)^D$, the effective extension of a DLT with DLT
boundary conditions in all directions is
\begin{equation} \label{Nseff}
   N_s^{\rm eff} = 2^{1/D}\,N_s\,,
\end{equation}
so that $(N_s^{\rm eff})^D$ is the size of the DLT. One may argue that 
finite length corrections are exponentially suppressed by $\sqrt{2}\,
N_s$, which is for $D>2$ larger than $N_s^{\rm eff}$. Then one would 
for $D\ge 3$ gain with respect to the suppression of finite size effect 
compared to the usual torus.  However, simulations of the 3D and 4D 
Ising model on a DLT \cite{BeHe07} showed an exponential suppression 
of finite size corrections with $N_s^{\rm eff}$ and not with $\sqrt{2}
\,N_s$. The reason for that has remained unclear.

When using two different $\beta$ values, $\beta_0\ne\beta_1$, we 
assign a unique $\beta_i$, $i=0,1$ to each plaquette in a slightly 
asymmetrical way: If any link of a plaquette is from the second torus, 
we take $\beta_1$, otherwise $\beta_0$. Technically this is done by 
tagging all links in the first torus by $0$ and in the second by~$1$. 
When considering a plaquette, all these tags are added up. If the sum 
is zero, $\beta_0$ is used, otherwise $\beta_1$. So the $\beta_1$ 
lattice becomes slightly larger than the $\beta_0$ lattice.

\subsection{Parameter files}

As indicated in Fig.~\ref{fig_STMC2LSU3} runs are kept in subfolders of 
project folders, one run per subfolder. The relevant parameters are set 
in two files: {\tt latmpi.par} and {\tt mc.par}. Before the compile step 
the parameters are transferred by a simple preprocessing procedure into 
subroutines and, in particular, used to dimension common blocks properly 
(see section~\ref{subsec_code}). Due to this procedure it is {\it 
mandatory} that runs and their parameter files are kept two levels down 
from the {\tt STMC2LSU3MPI} root directory.

As an example the parameter files of the run in
$$ {\tt 1MPICH/08x08y08z04t5p65b2f3d} $$
are given below. \smallskip

\centerline{\tt latmpi.par} \smallskip

\begin{tiny} \begin{verbatim}
c Bernd Berg Jan 11, 2009. MPI Checkerboard lattice:
c cfln      Part of data file name.
c lbcex     Boundary exchange T/F.
c lsd2mpi   .true. distinct random number seeds on each process,
c           .false. identical random number seeds (for tests).
c nd        Dimension of the lattice space
c nl1       Lattice extension in first directions. Also used as 
c           label for the entire lattice. (nl2, nl3, nl4 lattices 
c           extension in directions 2, 3 and 4).
c nltau     Disfunctional. Planned option for nlat=2 to subdivide 
c           the nl4 direction of one torus (gives higher T). 
c           Presently not implemented: nltau=nl4 required.
c mxs       Number of space sites, mxsc same per checkerboard.
c ms        Number of sites, msc per checkerboard.
c mlink     Number of links, mlinkc links per checkerboard.
c mp        Number of plaquettes (mpc plaquettes per checkerboard).
      character cfln*6,cfile*18
      parameter(cfln="SU3LGT",lbcex=.true.,lsd2mpi=.true.)
      parameter(nd=4,ndm1=nd-1,n2d=2*nd,nl1=4,nl2=4,nl3=4,nl4=4)
      parameter(nltau=nl4) ! Always (purpose not implemented).
      parameter(mxs=nl1*nl2*nl3,mxsc=mxs/2,ms=mxs*nl4,msc=ms/2)
      parameter(mlink=nd*ms,mlinkc=nd*msc)
      parameter(mp=ms*(nd*(nd-1))/2,mpc=mp/2)
c nptime    Number of timelike plaquettes. 
c npspace   Number of spacelike plaquettes. 
      parameter(mptime=ms*(nd-1),mpspace=mp-mptime)
c nlat      1 or 2 layers (not other values allowed).
c lat2      must be false for nlat=1, active (false or true) only for 
c           nlat=2: .true. boundary exchange between DLT layers; 
c               .false. no boundary exchange between DLT layers.
c lat2test  .true. sets identical random numbers in both layers.
c ndmpi     Dimension of the MPI lattice.
c mpifactor Number of sublattices per ndmpi direction.
c msmpi     Total number of MPI sublattices (MPI processes) per layer.
c mpmpi     Total number of plaquettes over all sublattices (one layer).
c mscb      Size of one checkerbord boundary.
c mbcs      Extra storage size for boundaries, which 
c           enters definition of nmat below. 
c mbcsh     Extension of pointer array sizes to gather boundaries.
c noffset   For definition of receive position in 1. gather/scatter.
      parameter(nlat=1,lat2=.false.,lat2test=.false.)
      parameter(ndmpi=3,mpifactor=2,msmpi=mpifactor**ndmpi)
      parameter(n2dmpi=2*ndmpi,mpmpi=msmpi*mp)
      parameter(mscb=nl2*nl3*nl4/2,mbcs=n2dmpi*mscb)
      parameter(mbcsh=mbcs/2,noffset=msc+ndmpi*mscb)
c nddmpi    Number of combinations of two ndmpi directions.
c ns1fbb    Number of matrices for gather in 2. gather/scatter,
c           which is for corner plaquettes.
c nsfbb     For check on number of corner plaquetted in cblgtpnt2.f.
c           It enters definition of nmat below.
c nmat      Number of SU3 matrices including those gathered from 
c           neighbouring sublattices.
c n18       Number of SU3 matrix elements.
      parameter(nddmpi=ndmpi*(ndmpi-1),ns1fbb=nl3*nl4/2) 
      parameter(nsfbb=nddmpi*ns1fbb,nmat=msc+mbcs+nsfbb,n18=18)
C Array sizes in latmpi.dat have to be consistent!!!
\end{verbatim} \end{tiny} 
 \smallskip

Central are the sublattice extensions, {\tt nl1}, {\tt nl2}, {\tt nl3}, 
{\tt nl4} in 4D, and the MPI parameters {\tt ndmpi}, {\tt mpifactor}. 
For $\tt mpifactor=1$ there is only one process and the entire lattice 
agrees with the sublattice. For $\tt mpifactor>1$ the extensions of the
entire lattice agree with those of the sublattice in directions larger 
than {\tt ndmpi}, which exist for ${\tt ndmpi < nd}$, and there are {\tt 
mpifactor} sublattices in each of the {\tt ndmpi} directions. The 
sublattices themselves form a lattice of dimension {\tt ndmpi} with
\begin{equation} \label{MPIlat}
  \tt msmpi = mpifactor**ndmpi
\end{equation}
points, which we refer to as {\it MPI lattice}. To allow for variable
extensions, the sublattice values are stored in an array {\tt nla}.
To make use of the same routines, the MPI lattice extension {\tt 
mpifactor} is similarly stored in an array {\tt nla\_mpi}. Both arrays
are initialized in the file {\tt latmpi.dat}: 
\smallskip

\begin{small} \begin{verbatim}
   data nla/nl1,nl2,nl3,nl4/,nla_mpi/ndmpi*mpifactor/
\end{verbatim} \end{small} \smallskip

\noindent Usual PBC are simulated for $\tt nlat=1$, the DLT for $\tt 
nlat=2$. For PBC the number of MPI processes in {\tt msmpi}, while for 
the DLT it is {\tt 2*msmpi}. We will get familiar with choices of other 
{\tt latmpi.par} parameters when we perform verification and test runs 
in the next section.

The example file for {\tt mc.par} is: \smallskip

\begin{tiny} \begin{verbatim}
c Bernd Berg, Jan 11, 2009.
c Definition of parameters for SU(3) LGT simulations.
c Output units iuo, iud1, iud2 (off/on with lud2), iud3.
c Job number njob and seeds iseed1, iseed2.
      parameter(iuo=6,iud1=11,iud2=12,lud2=.false.,iud3=13)
      parameter(njob=1,iseed1=njob,iseed2=0)
c Parameters for equilibrium and data production sweeps:
c beta0,1: beta_g=2/g^2 inverse bare coupling.
c istart:  1 ordered start all matrices 1; 2 disordered start.
c nhit:    Number FHKP proposals made.
c nreq:    Number of repititions of nequi equilibrium sweeps.
c nequi:   Number of equilibrium sweeps.
c nrpt:    Number of repititions of nmeas measurement. 
c nsw:     Number of sweeps between measurements.
c nmeas:   Number of measurement sweeps per repitition.
      parameter(beta0=5.65d00,beta1=beta0, istart=1,nhit=1)
      parameter(nreq=1,nequi=2**12,nrpt=32,nsw=2,nmeas=nequi)
\end{verbatim} \end{tiny} 
 \smallskip

The purpose of most parameters should be obvious from the comments. 
The MCMC run structure is that defined by {\tt nequi}, {\tt nrpt} 
and {\tt nmeas} in Ref.~\cite{Be04}. After equilibration measurements 
are saved in {\tt nrpt} blocks to allow for a conveniently binned 
analysis, employing jackknife methods when suitable. There are
{\tt nsw} sweeps done between measurements.

\subsection{Program structure \label{subsec_code}}

We trace the code structure and that of a typical run from the main 
program 
\begin{equation} \label{SU3main}
  {\tt cbsu3\_dlt\{a,b,c\}.f}\ . 
\end{equation}
The program comes in three versions $\tt \{a,b,c\}$, where {\tt b} is 
obtained from {\tt a} by simply replacing everywhere in the code {\tt 
mpia} by {\tt mpib}, and similarly for {\tt c}. As discussed in 
\cite{BeWu09}, differences lie in the coding of MPI send and receive 
instructions. We were unable to find a single solution which works on 
all MPI platforms on which we performed tests. The {\tt a} version, 
which uses the simplest (plain) subroutines for boundary transfers,
is listed in the following. \smallskip

\input CODE/cbsu3_dlta.f \smallskip

In the first lines of the program, after the comments, the general
structure is defined. Variables are declared throughout the entire
code by including the {\tt implicit.08} file of the {\tt Fortran}
library folder: \smallskip

\begin{verbatim}
      implicit real*8 (a-h,o-z)
      implicit logical (l)
\end{verbatim} 
\smallskip

\noindent This has the advantage that the type of a variable follows
from its first letter. An exception to this rule are character
variables, which are explicitly declared, though their first letter
is always {\tt c}. No complex variables are used. MPI is setup by 
including the system provided file {\tt mpif.h} and a number of 
constants are defined by including the file {\tt constants.08} 
(see inside the file).

The program is compiled by a file {\tt mpimake}, or similar, of which
a copy is located in each project folder and listed here as used 
for Open MPI\footnote{Similar {\tt mpichmake} files are included in 
the MPICH  folders. For runs on the Cray the compile step is in the 
job {\tt q.run*} control cards, because of the queuing system there. 
In our Open MPI installation we had to use the {\tt b} version of the 
program, {\tt c} on the Cray, while the {\tt a} version is sufficient 
for some of the runs we performed on the Cray and all the MPICH runs 
documented in this paper.}. \smallskip

\begin{verbatim}
cp *.par ../../Libs/Fortran_par/.
mpif77 -O -Wall $1
rm ../../Libs/Fortran_par/*.par
\end{verbatim} 

\smallskip \noindent
The {\tt mpimake} command transfers the parameter files into the {\tt 
Fortran\_par} folder and removes them from there after the compile 
step. This creates a hyperstructure, which transfers to all subroutines 
identical parameter values and dimensions common blocks properly. As 
already mentioned, to keep this structure intact runs {\it must} 
be carried out in subfolders, which are two levels down from {\tt 
STMC2LSU3MPI}. Job submission is subsequently done by {\tt run*} 
executables, which are kept in the run subfolders.

All library routines needed by the program are explicitly included at 
the end of the main program. So their source code can be easily located. 
Exceptions are calls to MPI routines (all routines with names starting 
with {\tt mpi}), which have to be looked up in MPI manuals or tutorials
(for instance~\cite{Be04}, see \cite{BeWu09} for subtle points with 
send and receive). Step by step the execution of a run is explained 
in the following. 

\begin{figure}[t]
 \begin{picture}(150,155)
    \put(0, 0){\includegraphics{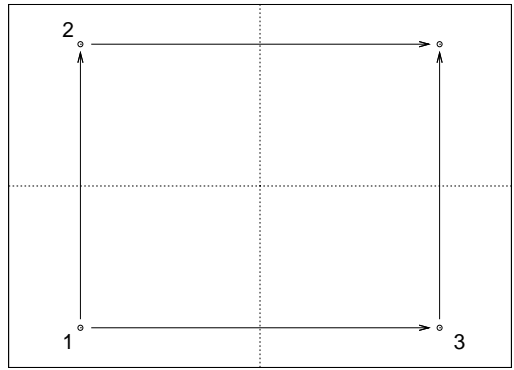}}
  \end{picture}
\caption{Links emerging at sites~2 and~3 are needed for updates of 
links emerging from site~1 (the broken lines indicate a division
into sublattices). \label{fig_BCcorner}}
\end{figure}

\begin{enumerate}
\item MPI initialization by {\tt mpi\_init}.
\item Calculation of the rank (identity {\tt my\_id} of the MPI 
      process) by {\tt mpi\_comm\_rank}.
\item Some printout from MPI process zero, setup of printout for 
      each process if {\tt lud2} is true.
\item A call to {\tt cbsu3\_2init\_mpi} initializes the run, setting
      up many important features:
      \begin{enumerate}
      \item A call to {\tt rmaset} initializes Marsaglia's (pseudo) 
            random number generators \cite{Ma90,Be04} used throughout 
            this code. For {\tt lsd2mpi} true the process rank is 
            invoked in the seed, so that a different generator is 
            used for each MPI process.
      \item Definition of pointer arrays for checkerboard labeling and 
            exchange of boundaries by calls to {\tt lat\_init}, for {\tt 
            nlat=2} also to {\tt lat2a\_init} and {\tt lat2b\_init}, 
            then to {\tt cblat\_init}, for {\tt lbcex} true (means MPI 
            boundary conditions exchange) to {\tt cblgtpointer} and, 
            finally, for {\tt ndmpi}$\ge$2 to {\tt cblgtpnt2}. 
      \item For the DLT a call to {\tt cbsu3\_iba\_mpi} assigns a 
            unique $\beta$ to each plaquette. This routine has to be 
            called before the start configuration is initialized, 
            because it uses the SU(3) matrix array for temporary 
            storage.  Tags are finally stored in the arrays {\tt iba1} 
            and {\tt iab2} of the common block {\tt common\_cbsu3.f} 
            and pointers to the $\beta$ values in the array 
            {\tt ba(0:4)}.
      \item A call to {\tt csu3\_start} generates a SU(3) start
            configuration.
      \item A call to {\tt cbsu3\_act\_mpi} calculates the initial
            action.
      \end{enumerate}
\item Calls to {\tt cbsu3\_actdif\_mpi} check whether the action
      kept on record during the updating process agrees with the one 
      obtained by direct calculation. Process~0 writes action 
      information to the formatted output file (unit {\tt iuo}) 
      through calls to {\tt write\_act\_mpi}.
\item Calls to {\tt mpi\_barrier} are supposed to synchronize the 
      MPI processes, but may indeed have no effect.
\item Calls to {\tt write\_progress} by MPI process~0 write 
      information to a file {\tt progress.d}, which is opened and 
      closed, so that the user can look up the file during run time.
\item For equilibration a double loop ({\tt nreq} and {\tt nequi}) 
      of calls to the updating routine {\tt cbsu3\_2hbnhit\_mpi} is 
      performed. The purpose of a double loop is that the run can 
      be interrupted when the total equilibration time exceeds the
      CPU time allowed for a single run. The updating routine relies 
      on a number of subroutines:
      \begin{enumerate}
      \item {\tt cbsu3\_bstaple1} calculates the staple for updating 
            a link matrix on checkerboard~1 ({\tt cbsu3\_bstaple2} 
            correspondingly on checkerboard~2). These routines use 
            various matrix manipulation routines from {\tt Libs/SU3}.
      \item {\tt su3mult\_m\_m\_m} multiplies SU(3) matrices of
            the first two arguments and returns the result in the 
            third argument.
      \item {\tt su3reunit} reunitarizes a SU(3) matrix.
      \item {\tt cbsu3\_bnd1a\_mpia} collects boundaries (no corners)
            from a sublattice checkerboard~1 and sends them to other 
            sublattices ({\tt cbsu3\_bnd2a\_mpia} for collection
            from checkerboard~2). A subtle point is in gauge systems 
            that one needs for {\tt ndmpi>1} corner links from two 
            neighboring sublattices like the links emerging from 
            sites~2 and~3 in Fig.~\ref{fig_BCcorner}. This is handled 
            by one more routine:
      \item {\tt cbsu3\_bnd1b\_mpia} collects for {\tt ndmpi>1}
            boundary corners from checkerboard~1 and sends them 
            to other sublattices ({\tt cbsu3\_bnd2b\_mpia} for 
            collection from checkerboard~2).
      \end{enumerate}
\item Updating sweeps with measurements are carried out in a triple loop 
      ({\tt nrpt}, {\tt nmeas} and {\tt nsw}). Measurements are done 
      every {\tt nsw} sweeps and kept in time series arrays of length 
      {\tt nmeas}. To write reasonably sized unformatted arrays to disk
      is considerably faster than writing after each measurement step. 
      Increasing {\tt nsw} prevents strongly correlated measurements. 
      A good choice for {\tt nsw} is between 1\% and 10\% of the 
      expected integrated autocorrelation time 
      $\tau_{\rm int}$~\cite{Be04}, which depends 
      not only on $\beta$ and the lattice size, but also on the 
      observable. Using a too large value for {\tt nsw} destroys the 
      possibility to estimate $\tau_{\rm int}$ from the run data.
\item Measurements are temporarily stored in arrays of the common block 
      {\tt common\_cbsu3}. For spacelike and timelike plaquettes 
      they are done by {\tt cbsu3\_wloops\_mpi} and kept in the 
      times series (ts) arrays {\tt tsws} and {\tt tswt}. 
\end{enumerate}

\section{Verifications} \label{Verify}

Although our code is written for a variable lattice dimension D, tests 
have so far been limited to 4D. The programs and routines are only 
moderately cleaned up. Many parts have disabled ({\tt ltest=.false.})
or commented out test options. They are presently left in the code, 
because they could come into use again. 

This section deals with verifications, which were performed on a 2~GHz 
AMD Athlon 64 XM Dual Core Processor 3600+ at Leipzig University. MPI 
runs with ${\tt mpifactor}>1$ use both processors, single processor 
runs one of them. Fortran~77 compilation was done with the g77 compiler 
based on gcc version 4.3.2 (Debian 4.3.2-1.1). MPI runs were performed 
with MPICH version 1.27p1. Compiler warnings about slow initialization 
of large aggregate areas have been ignored as the produced code works 
just fine. More MPI runs using up to 16 cores on a PC cluster and up 
to 1$\,$296 on a Cray are documented in~\cite{BeWu09}.

A strong test for the correct implementation of exchange of boundaries 
is provided by using identical random numbers on each sublattice. Then 
results from all sublattices have to agree and be identical with a run 
on a single lattice of this size with PBC. After such tests, real 
production runs were performed to compare action expectation values 
with results from the literature \cite{Bo96} and from our conventional 
(non-MPI) SU(3) code used before in Ref.~\cite{BaBe07} (they are setup 
in the {\tt STMCSU3} project folder in essentially the same way as the 
MPI programs in the main tree).

Because there are no IEEE (Institute for Electrical and Electronics 
Engineers) standards for Fortran functions, the precise numbers 
obtained in trial runs depend on the computing platform due to 
rounding errors, which lead at some point to distinct accept/reject 
steps. For averages agreement in the statistical sense has to hold. 
This is still very restrictive as the statistical errors are often 
small.

\subsection{Periodic boundary conditions} \label{Verify1}

This section deals with verifications for simulations with PBC, i.e., 
the parameter values 
$$  \tt nlat = 1~~and~~lat2=.false.\ . $$
All parameters of specific runs are kept in the subfolders of the 
project
$$\tt 1MPICH\ . $$

\subsection{Identical random numbers on sublattices}

With the
$$\tt lsd2mpi = .false. $$
option identical random numbers are used in all sublattices. We 
performed such simulations on $4^4$ sublattices with parameters
\begin{equation} \label{1STAT}
 \tt nmeas = nequi = 2^{12},\ nrpt = 32,\ nsw = 2
\end{equation}
at $\beta=5.5$, 5.6 and~5.7 using ordered and disordered starts. The 
average actions of the MPI runs are obtained by running the analysis 
program
\begin{equation} \label{ans2}
 \tt ana2sublatsu3.f
\end{equation}
and the values were found in statistical agreement with those from a 
conventional single processor SU(3) Fortran program: The average over 
the $Q$ values of Gaussian difference tests \cite{Be04} between these 
six runs was
close to 0.5 as it should. 

We document here only the 
$$ \beta=5.6~~{\rm runs\ with\ ordered\ starts}\ .$$
The analysis is kept on {\tt ana2.txt} files. We obtained for the mean 
action per plaquette with MPI code (error bars are given in parenthesis 
and always rounded upwards in their second digit)
\begin{equation} \label{act1MPI}
  {\tt act} = 0.53811\, (19)\,,
\end{equation}
versus with single processor code
\begin{equation} \label{act1noMPI}
  {\tt act} = 0.53770\, (18)\,, 
\end{equation}
leading to an acceptable $Q=0.12$ in the Gaussian difference test. 
The integrated autocorrelation time of these runs is estimated to 
be $\tau_{\rm int} = 49.5\, (3.4)$. So an error bar calculation with 
respect to 32 bins is appropriate.

MPI runs were repeated for the pairs (1,1), (2,1), (2,2), (2,3), 
(2,4), (3,1), (3,2) of the parameters
\begin{equation} \label{pairs}
  \tt (mpifactor,ndmpi)
\end{equation}
giving (due to {\tt lsd2mpi} false) always to the same average 
action~(\ref{act1MPI}). The corresponding numbers of MPI lattice points 
(MPI processes) {\tt msmpi} (\ref{MPIlat}) are 1, 2, 4, 8, 16, 3 and 9. 

Parameters of the runs of this section are kept in
\begin{equation} \label{LatNotation}
  \tt \{F\}nnxnnynnznntnpnbnfnd
\end{equation}
subfolders of {\tt 1MPICH}, where {\tt F} indicates $\tt lsd2mpi= 
.false.$ and is omitted for $\tt lsd2mpi=.true.$. The letters {\tt n}
indicate numbers, which can be different.  Lattice extensions are given
by {\tt nnx}, {\tt nny}, {\tt nnz} and {\tt nnt}. This is followed by 
{\tt npnb} for $\beta_0=\beta_1$={\tt n.n}, by {\tt n} from {\tt nf} 
for $\tt mpifactor=n$, and {\tt n} from {\tt nd} for $\tt ndmpi=n$. In 
the folder names extensions of the full lattice are used, whereas data 
set names created by the program (\ref{SU3main}) are of the form
\begin{equation} \label{DatNotation}
  \tt SU3LGTndnfndnnnxnnntnnnn.D\ ,
\end{equation}
showing {\it sublattice extensions} of the {\tt x} and {\tt t} 
directions. The program calculates also the extensions of the full 
lattice from {\tt latmpi.par} and prints them in the readable output 
file. Another way to find sublattice extensions is from the folder name 
by dividing the full lattice extensions by {\tt mpifactor} for the {\tt 
ndmpi} directions. The results have to be integers without rest term. 
As folder names are created by hand, the output file from the run is 
authoritative in case of a discrepancy. 

The other acronyms in the data set name (\ref{DatNotation}) are the 
lattice dimension ${\rm D}={\tt n}$ of the first {\tt nd}, then $\tt 
mpifactor=n$ from {\tt nf}, and the MPI lattice dimension $\tt ndmpi=n$ 
from the second {\tt nd}. The extension {\tt nnnn.D} labels data files 
by their process number. Each of the created data files corresponds 
to one of the sublattices. After data production, the 1-processor MPI 
program
\begin{equation} \label{datcollect}
  \tt su3datcollect.f
\end{equation}
condenses these data files into a single one for which the extensions 
{\tt tnnnn.D} are reduced to {\tt t.D}. When disk space fills up it 
is sufficient to keep only the {\tt *t.D} files.

To give an example, the subfolder name
$$ \tt F08x080y40z04t5p6b2f2d $$
corresponds to ${\tt lsd2mpi=.false.}$ runs on $4^4$ sublattices at 
$\beta=5.6$ with a full lattice size $8^2\,4^2$. The MPI run produces 
four sublattice data sets
$$ \tt SU3LGT4d2f2d004x004tnnnn.D $$
with {\tt nnnn} from {\tt 0000} to {\tt 0003}. After data collection 
with (\ref{datcollect}) the file
$$ \tt SU3LGT4d2f2d004x004t.D $$
results, which can be analyzed further.

\subsection{Different random numbers on sublattices}

\begin{table}[th] 
\centering
\caption{\label{tab01}{Runs with np MPI processes, {\tt mpifactor}=nf, 
{\tt ndmpi}=n on a periodic $8^3\,4$ lattice at $\beta=5.65$.}}
\medskip
\begin{tabular}{|c|c|c|c|c|c|c|c|c|c|}
\hline
np &nf &n&{\tt nl1}&{\tt nl2}&{\tt nl3}&{\tt nl4}
                    &{\tt time}& {\tt actm}    & $Q$  \\ \hline
$-$&$-$&$-$& 8& 8& 8& 4& 248 m & 0.538547 (70) & $-$  \\ \hline
 1 & 1F&   & 8& 8& 8& 4& 282 m & 0.538471 (61) & 0.41 \\ \hline
 1 & 1 &   & 8& 8& 8& 4& 287 m & 0.538471 (61) & 1.00 \\ \hline
 2 & 2 & 1 & 4& 8& 8& 4& 147 m & 0.538584 (71) & 0.23 \\ \hline
 4 & 2 & 2 & 4& 4& 8& 4& 150 m & 0.538562 (63) & 0.82 \\ \hline
 8 & 2 & 3 & 4& 4& 4& 4& 158 m & 0.538468 (81) & 0.36 \\ \hline
\end{tabular}
\end{table}

We set
$$\tt lsd2mpi = .true.~~and~~mpifactor = 2 $$
in {\tt latmpi.par} and produce data at $\beta =5.65$ from simulations 
of $8^3\,4$ lattices, which are partitioned into different numbers of 
sublattices. Average action densities are compiled in table~\ref{tab01}. 
The statistics of each run is the same as before (\ref{1STAT}). Each 
$Q$ value in the table corresponds to the Gaussian difference test with 
the action value in the row above. The number of MPI processes agrees 
with the number of sublattices given by (\ref{MPIlat}). The {\tt time} 
column contains the CPU time measured on the Athlon PC. The non-MPI 
run for the first row uses the FHKP heatbath in the repeat until 
accepted version. Whether it is slower of faster than the MPI program 
run for one process $\rm (nf=1)$ depends on the Fortran compiler and 
the MPI installation. Here it turns out to be faster, but that is not 
the case on the PCs used in \cite{BeWu09}. For $\rm nf=1$ one can turn 
off the boundary exchange in the MPI program by setting $\tt lbcex=
.false.$, indicated by 1F in the nf column. Results stay number by 
number identical and the small speedup is negligible for practical 
purposes, indicating that MPI send and receive is very efficient as 
long as communication stays within the same computer node. The decrease 
of CPU time from $\tt msmpi=1$ to $\tt mspi=2$ reflects the gain in 
real time due to using both cores of the PC. It is given by a factor 
slightly larger than 1/2 due to communication overhead. The parameters 
of these runs are setup in the
$$\tt 08x08y08z04t5p65bnfnd $$
subfolders of {\tt 1MPICH}, where the notation is as introduced in 
(\ref{LatNotation}). 

\begin{table}[th] 
\centering
\caption{\label{tab02}{Spacelike and timelike plaquette expectation 
values for comparison with Ref.~\cite{Bo96} ({\tt B} in the {\tt np} 
column): Runs on a $16^3\,4$ lattice at $\beta=5.65$ using our non-MPI 
program and MPI code with {\tt np} processes (sublattices). }}
\medskip
\begin{tabular}{|c|c|c|c|c|} \hline
{\tt np}& spacelike   & $Q$& timelike    & $Q$\\ \hline
     B  &0.537638 (17)& $-$&0.537692 (19)& $-$\\ 
    $-$ &0.537650 (15)&0.60&0.537711 (14)&0.42\\ 
     1  &0.537647 (17)&0.89&0.537701 (17)&0.65\\ 
     2  &0.537650 (16)&0.90&0.537704 (17)&0.90\\ 
     4  &0.537648 (16)&0.93&0.537708 (16)&0.86\\ 
     8  &0.537661 (18)&0.59&0.537714 (17)&0.80\\ \hline
\end{tabular} \end{table}

Results from $16^3\,4$ lattices, which allow for comparison with 
spacelike and timelike plaquette averages of the literature \cite{Bo96}, 
are compiled in table~\ref{tab02}. For our runs the statistics 
(\ref{1STAT}) was used again. The estimates of \cite{Bo96} rely on 
20$\,$000 to 40$\,$000 sweeps after thermalization. After collection
of our data by running (\ref{datcollect}), the analysis program 
$$\tt anaw\_cbsu3.f $$
estimates the expectation values for spacelike and timelike plaquettes.
Again, $Q$ values correspond to Gaussian difference tests with the row 
above. The slight increase of all values with increasing numbers of
processes is accidental and not reproduced when using different 
random number generator seeds.

\subsection{Double-layered torus} \label{Verify2}

This section deals with verifications for simulations with
$$\tt nlat = 2 $$
and folders of the runs are kept in 
$$\tt 2MPICH\ .$$
For $\tt lat2=.false.$ the exchange of boundaries is turned off and
one performs independent runs on two lattices with PBC. This is not 
very interesting and we discuss only runs with $\tt lat2=.true.$.

With exchange of boundaries turned on, a strong verification test is 
to run with different random numbers on the sublattices of a torus, 
but identical $\beta$ values and random numbers on each torus. These 
are the parameter options
$$\tt lsd2mpi = .true.~~and~~lat2test=.true.\ ,$$
designed to reproduce the action values of the run with PBC for which
statistics and sublattices match. For tori of size $8^34$ and $\beta_0
=\beta_1= 5.65$ these are values of table~\ref{tab01}. These test runs 
worked out as required and are setup in the 
$$ \tt T08x08y08z04t5p65bnfnd $$
folders of {\tt 2MPICH}, where the initial {\tt T} indicates that 
{\tt lat2test} is set to true.

\begin{table}[th] 
\centering
\caption{\label{tab03}{Runs with np MPI processes, {\tt mpifactor}=nf, 
{\tt ndmpi}=n on a double-layered $8^3\,4$ lattice at $\beta=5.55$, 
disordered starts. }} \medskip
\begin{tabular}{|c|c|c|c|c|c|c|c|c|} \hline
np &nf &n &{\tt nl1}&{\tt nl2}&{\tt nl3}&{\tt nl4}
                       &  {\tt actm}   & $Q$  \\ \hline
 2 & 1 & 1 & 8& 8& 8& 4& 0.510608 (33) & $-$  \\ \hline
 4 & 2 & 1 & 4& 8& 8& 4& 0.510671 (28) & 0.15 \\ \hline
 8 & 2 & 2 & 4& 4& 8& 4& 0.510597 (24) & 0.04 \\ \hline
16 & 2 & 3 & 4& 4& 4& 4& 0.510676 (32) & 0.05 \\ \hline
\end{tabular} \end{table}

Proper simulations on a DLT are performed with {\tt lat2test=.false.}.
Table~\ref{tab03} gives reference values for the action at $\beta_0
=\beta_1=5.55$ (at $\beta=5.65$ one needs more statistics due to
autocorrelations that become important for small error bars). The 
$Q$ values refer to Gaussian difference tests as in previous cases.
They are a bit on the low side, but it is clear that simply reordering
the comparison of data would change that. Also runs with a different 
compiler gave (in the same order as listed in the table) $Q=0.97$, 
0.09, and~0.30.

\section{Summary and Conclusions} \label{Conclusions}

The code of this paper allows Markov chain Monte Carlo calculations
of pure SU(3) lattice gauge theory on computers which have MPI 
installed. Besides the usual periodic boundary conditions, the geometry 
of a double-layered torus is implemented, which allows for distinct 
inside and outside environments (the ``inside'' of one lattice is 
the other's ``outside'' and vice versa). A considerable number of 
non-trivial verification are included in this paper. The CPU time 
performance of the code as function of the number of processors (more 
precisely CPU cores) is documented in a companion paper~\cite{BeWu09}.

Should the {\tt a}-version of the program ``hang-up'' without producing 
an error message, the cause are likely MPI send and receive problems 
which are discussed and partially resolved in~\cite{BeWu09}. Although 
designed for arbitrary D$\ge$2 dimensions, the code has presently only 
been tested in 4D. Hence, it is unlikely to work straightaway for other 
D values ({\tt nd} in {\tt lat.par}), though required fix-ups are 
expected to be minor. Of course, it is in the responsibility of the 
final user to perform stringent tests before applying the provided 
code to any purpose. 

\bigskip
\noindent {\bf Acknowledgments:}
The non-MPI part of this code is based on joint work with Alexei 
Bazavov~\cite{BaBe07}. Contributions are indicated in the code. 
I would like to thank Alexei Bazavov, Urs Heller and Wolfhard 
Janke for useful discussions.  This work was in part supported by the 
U.S.  Department of Energy under contract DE-FG02-97ER41022 and by the 
German Humboldt Foundation. I am indebted to Wolfhard Janke, Elmar
Bittner and other members of the Institute for Theoretical Physics 
of Leipzig University for their kind hospitality.

\end{document}